\documentstyle[12pt]{article}
\def\be{\begin{equation}}
\def\ee{\end{equation}}
\def\ba{\begin{eqnarray}}
\def\ea{\end{eqnarray}}
\def\L{{\cal L}_Y}
\def\S{{\cal S}}
\def\M{{\cal M}}
\newcommand{\dfrac}[2]{{\displaystyle \frac{#1}{#2}}}
\title{\bf Relating the scalar flavour-changing neutral couplings
to the CKM matrix}
\author{G.\ C.\ Branco,
W.\ Grimus\thanks{On leave of absence
(from October 1, 1995 to January 31, 1996)
from the University of Vienna,
Institute for Theoretical Physics,
Boltzmanngasse 5, A-1090 Vienna}\ \
and L.\ Lavoura\thanks{Researcher of the
Technical University of Lisbon;
e-mail address i009@beta.ist.utl.pt}  \\
\\
\small CFIF, Instituto Superior T\'ecnico,
Edif\'\i cio Ci\^encia (f\'\i sica) \\
\small P-1096 Lisboa Codex, Portugal}
\begin{document}
\maketitle
\begin{abstract}
We build a class of two-Higgs-doublet models
in which the flavour-changing couplings of the neutral scalars
are related in an exact way to elements of the quark mixing matrix.
In this framework,
we explore the different possibilities for CP violation
and find some interesting scenarios,
like a realization of the superweak idea
without CP violation in the $B$-meson system.
In another scenario the neutral scalars can be relatively light,
and their contributions to the $B^0$--$\bar{B}^0$ transitions
can alter the standard-model
predictions for CP violation in that system.
\end{abstract}
%
%
\section{Introduction}
The introduction of more than one Higgs doublet in the standard model (SM)
leads in general to flavour-changing neutral interactions (FCNI)
at the tree level,
mediated by the physical neutral scalars.
In a general multi-Higgs-doublet model,
the flavour-changing couplings are unknown parameters.
However,
some of the FCNI are tightly constrained
by the smallness of the $K^0$--$\bar{K}^0$
and $B^0$--$\bar{B}^0$ mass differences,
and of the CP-violating parameter $\epsilon$ of the neutral-kaon system.
As a result,
if one does not assume any special suppression
of the flavour-changing neutral couplings,
neutral scalars must be very heavy \cite{lee}
with masses of order of at least a few TeV.

The purpose of this paper is
to put forward a class of two-Higgs-doublet models
in which the flavour-changing couplings are related in an exact way
(i.e.,
without any {\it ad hoc} assumptions)
to matrix elements of the Cabibbo--Kobayashi--Maskawa (CKM) matrix $V$.
This will be achieved through the introduction in the Lagrangian
of an exact global symmetry $\S$,
which may be either softly or spontaneously broken.
As a consequence of that symmetry,
there are no FCNI in the up-type-quark sector
(i.e.,
among the quarks with electric charge $2/3$),
while the flavour-changing neutral vertex between two down-type quarks
$i$ and $j$ is suppressed by a factor $V_{\alpha i}^\ast V_{\alpha j}$,
where $\alpha$ denotes one of the up-type quarks.
The different models in our class of models
are distinguished by the following features.
First,
the quark $\alpha$ may be either the up,
the charm or the top,
and then the suppression factors in the various FCNI are of course different.
Second,
CP violation in our models may be either hard (HCPV) or spontaneous (SCPV),
depending on the Higgs potential
and leading to different predictions of the models.

Some of the models that we suggest have quite interesting features.
For instance,
in one of them,
in which CP violation is hard and the quark $\alpha$ is the top,
the flavour-changing interactions between the down and the strange quarks
are strongly suppressed,
and as a consequence the neutral-Higgs contributions
to the $K^0$--$\bar{K}^0$ mass difference and to $\epsilon$ are negligible;
on the other hand,
their contribution to the $B^0$--$\bar{B}^0$ transition-matrix element
may be substantial,
and have a phase different from the SM one.
They may thus affect the predictions for CP violation in the $B$-meson systems.
In that model,
the mass of the neutral scalars may be quite low,
of order a hundred GeV.
In another model,
$\alpha$ is the up or charm quark,
and CP violation is spontaneous.
The inverse situation then occurs:
the neutral scalars must be rather heavy,
$\epsilon$ originates entirely in the tree-level scalar-exchange diagram,
and the model has a superweak character \cite{superweak},
with the peculiar feature that,
because the FCNI in the $B$-meson systems are very much suppressed,
CP violation in those systems is predicted to be virtually unobservable.

The interest in the question of the suppression of the FCNI
has recently been revived \cite{hall} by the suggestion
that this suppression may be provided by family symmetries.
Some authors \cite{geo,lav}
have previously done work which relates to ours
in that they have introduced a horizontal symmetry
which has a similar structure to the one we employ.
Also,
it has already been suggested \cite{rin,josh}
that the FCNI among the down quarks may be suppressed by factors
$V_{\alpha i}^\ast V_{\alpha j}$,
just as we do it here;
however,
a horizontal symmetry leading to that suggestion has not been indicated.

Our paper is organized as follows.
In section 2 we discuss the symmetry ${\cal S}$ and the Yukawa couplings
in our class of models,
and derive the suppression factors of the FCNI.
In section 3 we focus on the Higgs sectors and Higgs potentials
resulting in different types of CP violation.
In section 4 we consider the constraints arising
from the $K^0$--$\bar{K}^0$ and $B^0$--$\bar{B}^0$ transitions,
and apply them to our models.
In section 5 we emphasize and summarize the main points of our work.
\section{The Yukawa couplings}
In order to fix our notation
we write down the most general Yukawa Lagrangian
for the quark multiplets of the SM and two Higgs doublets,
which is
\be \label{L}
\L = - \sum_{j=1}^{2} \left[
\bar{L} \Gamma_j \Phi_j n_R
+ \bar{L} \Delta_j
(i \sigma_2 \Phi_j^*) p_R
\right] + h.c.,
\ee
where $L = (p_L, n_L)^T$
denotes the 3-vector of left-handed doublets and $p_R$,
$n_R$ the right-handed singlets with hypercharge $4/3$ and $-2/3$,
respectively.
After spontaneous symmetry breaking,
the vacuum expectation values (VEVs)
\be \label{vacuum}
<\Phi_j>_0 = \frac{v_j e^{i \alpha_j}}{\sqrt{2}} 
\left( \begin{array}{c} 0 \\ 1 \end{array} \right) , \quad j =1,2
\ee
with $v_1 > 0$ and $v_2 > 0$ generate the quark mass matrices
\be
M_n = \frac{1}{\sqrt{2}} ( v_1 e^{i \alpha_1} \Gamma_1 + 
                           v_2 e^{i \alpha_2} \Gamma_2 ), \quad
M_p = \frac{1}{\sqrt{2}} ( v_1 e^{-i \alpha_1} \Delta_1 + 
                           v_2 e^{-i \alpha_2} \Delta_2 ),
\ee
which are diagonalized by the basis transformations
\be
n_L = U_L^d d_L, \; n_R = U_R^d d_R, \; p_L = U_L^u u_L, \; p_R = U_R^u u_R,
\ee
with mass eigenstates $d_L$,
$d_R$,
$u_L$,
$u_R$ and
\be
D_d = (U_L^d)^\dagger M_n U_R^d, \quad
D_u = (U_L^u)^\dagger M_p U_R^u
\ee
being the diagonal mass matrices for the down and up sectors,
respectively.
The CKM matrix is $V = (U_L^u)^\dagger U_L^d$.

It is useful to make the decomposition
\be
\Phi_j = e^{i \alpha_j}
\left( \begin{array}{c}
\phi_j^+ \\
\frac{1}{\sqrt{2}} (v_j + \rho_j + i \eta_j)
\end{array} \right) ,
\quad j=1,2
\ee
with real scalar fields $\rho_j$,
$\eta_j$.
The pseudo-Goldstone bosons $G^+$,
$G^0$ are singled out by the transformations
\be
\left( \begin{array}{c} G^+ \\ H^+ \end{array} \right) =
O \left( \begin{array}{c} \phi_1^+ \\ \phi_2^+ \end{array} \right) ,\;
\left( \begin{array}{c} G^0 \\ I \end{array} \right) =
O \left( \begin{array}{c} \eta_1 \\ \eta_2 \end{array} \right) ,\;
\left( \begin{array}{c} H^0 \\ R \end{array} \right) =
O \left( \begin{array}{c} \rho_1 \\ \rho_2 \end{array} \right) ,
\ee
with
\be
O = \frac{1}{v} \left( \begin{array}{lr} v_1 & v_2 \\ v_2 & -v_1 
                       \end{array} \right), \mbox{  where  }
v \equiv \sqrt{v_1^2 + v_2^2} = (\sqrt{2} G_F)^{-1/2} \approx \mbox{246 GeV}  
\ee
and $G_F$ is the Fermi coupling constant.
Since we do not want to introduce additional charged scalars,
$H^+$ is already the physical charged Higgs particle,
whereas in the case of the neutral scalars we take into account
the possibility of gauge singlets appearing in the Higgs potential.
Therefore,
we relate the physical neutral scalar mass eigenstates $\varphi_k$,
$k=1,2,3,\dots$ to $H^0$,
$R$,
$I,\dots$ via
\be
\left( \begin{array}{c} H^0 \\ R \\ I \\ \vdots \end{array} \right) =
\left( \begin{array}{c} W_{1k} \varphi_k \\ W_{2k} \varphi_k \\ 
                        W_{3k} \varphi_k \\ \vdots \end{array} \right)
\ee
with an orthogonal matrix $W$.

As announced in the introduction,
for the purpose of relating the flavour-changing neutral Higgs interactions
with CKM matrix elements,
we introduce the horizontal symmetry
\be \label{hor}
\S: \quad L_1 \rightarrow \omega L_1, \;
p_{R1} \rightarrow \omega^2 p_{R1}, \;
\Phi_2 \rightarrow \omega \Phi_2,
\ee
and all other fields transforming trivially under $\S$.
For the time being we do not fix $\omega$
but merely impose the restrictions $\omega \not= 1$ and $\omega^2 \not= 1$,
apart from the trivial $|\omega| =1$.
Possible values of $\omega$ will be determined by the Higgs potential.
With the above restrictions $\S$ leads to the Yukawa coupling matrices
\be \label{yu} \nonumber
\Gamma_1 = \left( \begin{array}{ccc} 
           0 & 0 & 0 \\ a_{21} & a_{22} & a_{23} \\ a_{31} & a_{32} & a_{33}
                  \end{array} \right) , \quad
\Gamma_2 = \left( \begin{array}{ccc} 
           b_{11} & b_{12} & b_{13} \\ 0 & 0 & 0 \\ 0 & 0 & 0 
                  \end{array} \right),
\ee
\be \label{yuc}
\Delta_1 = \left( \begin{array}{ccc}  
           0 & 0 & 0 \\ 0 & c_{22} & c_{23} \\ 0 & c_{32} & c_{33}
                  \end{array} \right) , \quad
\Delta_2 = \left( \begin{array}{ccc}  
           d_{11} & 0 & 0 \\ 0 & 0 & 0 \\ 0 & 0 & 0 

                  \end{array} \right).
\ee
The special form of these matrices yields
\be \label{S}
(U_L^d)_{1i} = V_{1i} \equiv S_{i},
\ee
because $U_L^u$ has the same block structure as $\Delta_1$ and $\Delta_2$
in eq.~(\ref{yuc}),
while $(U_L^u)_{11}$ can be chosen to be 1.

We have not specified with which of the up-type quarks the index 1
in eq.~(\ref{S}) is associated;
it can be associated with any of the up-type quarks and,
therefore,
in the following we will have to deal with three different 
models named after the up-type quarks u,
c,
and t.

Let us suppose that we want CP violation to be spontaneous,
and therefore we take the four matrices $\Gamma_j$ and $\Delta_j$ to be real.
The phases of the VEVs,
$\alpha_1$ and $\alpha_2$,
then feed into the Yukawa couplings,
but because of the form of those matrices they can be absorbed
by redefinitions of the phases of the $L_i$,
$p_{Ri}$ and $n_{Ri}$.
Then,
the CKM matrix is real,
i.e.,
in spite of the spontaneous CP violation both the CKM matrix
and the Yukawa coupling matrices in the mass basis (see below)
remain real.

It is easy to calculate the Yukawa couplings of the physical scalars:
\ba
\L (\varphi) & = & \frac{\varphi_k}{v}
\bar{d} \left[
- D_d W_{1k}
+ (N_d P_R + N_d^\dagger P_L ) W_{2k}
+ i (N_d P_R - N_d^\dagger P_L ) W_{3k}
\right] d
\nonumber \\
 & &
- \frac{\varphi_k}{v} \bar{u} \left(
D_u W_{1k}
+ N_u W_{2k}
- i \gamma_5 N_u W_{3k}
\right) u ,\label{neutral}\\
\L (H) & = &
\frac{\sqrt{2} H^+}{v} \bar{u} \left(
V N_d P_R + N_u V P_L
\right) d + \mbox{h.c.},
\label{charged} \ea
with $P_L = (1 - \gamma_5)/2$,
$P_R = (1 + \gamma_5)/2$ and
\ba \label{N}
N_d & = &
- \frac{v_2}{v_1} D_d +
\left( \frac{v_2}{v_1} + \frac{v_1}{v_2} \right)
S^\dagger S D_d,
\nonumber \\
N_u & = &
- \frac{v_1}{v_2} \mbox{diag} (m_{u1},0,0) +
\frac{v_2}{v_1} \mbox{diag} (0,m_{u2},m_{u3}) .
\ea
This equation makes it clear that in our class of models
there are no flavour-changing neutral interactions in the up sector,
whereas in the down sector they are related to the elements of the CKM matrix.
We want to stress the fact that
because of the horizontal symmetry in eq.~(\ref{hor}),
the tree-level results of eqs.~(\ref{N}) are exact.
This is in contrast to refs.~\cite{rin,josh},
where similar relations were obtained by choosing $U_L^u =$ {\bf 1}
without a justifying symmetry.
The symmetry $\S$ is essential to guarantee that the results
of eqs.~(\ref{N}) are stable under renormalization.

Our Yukawa coupling matrices in eqs.~(\ref{yu}), (\ref{yuc})
have the further property that
the phases of the VEVs $\alpha_1$ and $\alpha_2$
can be absorbed by redefinitions of the phases of $p_{L,R}$ and $n_{L,R}$.
As a consequence,
if the Yukawa coupling matrices $\Gamma_{1,2}$ and $\Delta_{1,2}$ are real,
there is no CP violation in the fermion sector,
i.e.,
the matrices $V$,
$N_d$ and $N_u$ are real.
\section{The Higgs potential}
As stated in eq.~(\ref{hor}),
the symmetry ${\cal S}$ acts on the Higgs doublets
in the following way:
$\Phi_1 \rightarrow \Phi_1$ and $\Phi_2 \rightarrow \omega \Phi_2$,
with $\omega \neq 1$ and $\omega^2 \neq 1$.
This raises the problem that
neither $\Phi_1^\dagger \Phi_2$
nor $(\Phi_1^\dagger \Phi_2)^2$
are invariant under ${\cal S}$,
and therefore there is an accidental symmetry in the Higgs potential leading,
upon spontaneous symmetry breaking,
to a pseudo-Goldstone boson.
The different models in our class of models
will be defined by the various ways of avoiding this accidental symmetry
and also by the CP properties of the Higgs potential.
The common feature of the class of models we consider
is the suppression of FCNI implied in eqs.~(\ref{N}).
As far as CP is concerned,
it is important to specify whether CP is spontaneously broken or not
and whether the CKM matrix has a complex phase or not.
In either case,
it is crucial to identify all sources of CP violation in a given model.
For example,
it is important to know whether there is CP violation via the mechanism
of scalar-pseudoscalar mixing \cite{lee,ma}.
In order to get that kind of CP violation,
we must have a Higgs potential with a phase structure rich enough
to generate CP violation in the Higgs sector alone,
independently of the presence of fermions in the model.
We should also consider the possibility of soft breaking of ${\cal S}$.
One may like to use it,
especially if one wants to have an argument of technical naturalness
to justify the smallness of some couplings.

All the above issues give us guidelines for the construction
of the Higgs potential.
In the following we consider four examples,
all of them realized with the two doublets $\Phi_1$ and $\Phi_2$
and additional gauge singlets of hypercharge zero.

\paragraph{Soft breaking of the accidental symmetry:}

We may eliminate the undesirable accidental symmetry
simply by introducing into the potential the term
$\mu \Phi_1^\dagger \Phi_2 + h.c.$,
which breaks both ${\cal S}$ and the accidental symmetry softly.
The Goldstone boson then acquires a squared mass
proportional to $|\mu|$.
Notice that $\mu$ may,
without loss of generality,
be taken to be real:
if it is complex,
its phase is compensated by the relative phase of the VEVs
$\alpha := \alpha_2 - \alpha_1$
(see eq.~(\ref{vacuum})),
and no scalar-pseudoscalar mixing arises.
Of course,
this is a consequence of the symmetry $\S$,
which eliminates term of the type $(\Phi_1^\dagger \Phi_2)^2$.

\paragraph{Elimination of the accidental symmetry:}

If one dislikes the soft breaking of symmetries,
the simplest alternative to get rid of the accidental symmetry
is assuming that $\omega = i$ and ${\cal S}$ generates a $Z_4$.
We introduce a complex gauge singlet $\sigma$,
transforming under ${\cal S}$ like $\sigma \rightarrow - i \sigma$.
Then,
the terms $(m \Phi_1^\dagger \Phi_2 \sigma + \lambda \sigma^4) + h.c.$
appear in the Higgs potential,
and they rid us of all accidental symmetries.
Once again,
if we assume that CP violation is hard then $m$ and $\lambda$ are complex,
but in spite of that the Higgs sector by itself is CP conserving.
Both $\alpha$ and the phase of the VEV of $\sigma$,
which we denote by $\beta$,
simply adjust themselves in order to cancel out
the phases of $m$ and $\lambda$.
No scalar-pseudoscalar mixing arises at tree level.\footnote{If $m$
and $\lambda$ are real positive,
the minimum of the above potential has $\alpha = \pi / 4$
and $\beta = 3 \pi / 4$ (or vice versa).
This minimum is CP-conserving.}

\paragraph{SCPV without soft breaking:}

It is often considered desirable to have CP violation to be spontaneous,
because SCPV bestows more predictive power on models.
In order to obtain SCPV without any soft-breaking terms,
the simplest alternative is the following.
We introduce two scalar singlets,
the complex $\sigma$ of the previous paragraph,
and a real $\chi$.
We let ${\cal S}$ be the same symmetry as in the previous paragraph,
and assume that $\chi \rightarrow - \chi$ under ${\cal S}$.
We then have four terms in the Higgs potential which can ``see'' phases:
$m (\Phi_1^\dagger \Phi_2 \sigma + h.c.)$ and
$\lambda (\sigma^4 + h.c.)$ as in the previous paragraph,
and also
$\lambda^\prime \chi (\Phi_1^\dagger \Phi_2 \sigma^\ast + h.c.)$ and
$m^\prime \chi (\sigma^2 + h.c.)$,
with real constants $m$,
$m^\prime$,
$\lambda$ and $\lambda^\prime$,
so that the potential is CP-invariant.
We obtain a vacuum potential where the phase dependence is contained
in the four terms $\cos(\alpha + \beta)$,
$\cos(\alpha - \beta)$,
$\cos(2 \beta)$ and $\cos(4 \beta)$.
This is enough to generate SCPV.

\paragraph{SCPV with soft breaking of ${\cal S}$:}

A simpler alternative for SCPV is possible if we allow for
soft breaking of the horizontal symmetry.
Once again we consider the same $Z_4$
and the same complex singlet as before,
but now we do not introduce any real singlet.
Besides the terms $m (\Phi_1^\dagger \Phi_2 \sigma + h.c.)$
and $\lambda (\sigma^4 + h.c.)$,
we now introduce into the Higgs potential the term
$\mu (\sigma^2 + h.c.)$,
which breaks ${\cal S}$ softly.
Then,
the $Z_2$ subgroup of $Z_4$ generated by ${\cal S}^2$,
under which $\Phi_2$ and $\sigma$ change sign,
while $\Phi_1$ remains invariant,
is preserved.
This $Z_2$ forbids any other soft-breaking terms
but $m^\prime (\Phi_1^\dagger \Phi_2 \sigma^\ast + h.c.)$;
however,
this term is of dimension 3,
higher than the dimension 2 of the original soft-breaking term
$\mu (\sigma^2 + h.c.)$,
and therefore we are allowed to neglect it
without destroying renormalizability.
The vacuum potential now has three terms depending on the phases
$\alpha$ and $\beta$:
\be \label{V0}
V_0 = \dots
+ a \cos(\alpha + \beta)
+ b \cos(4 \beta)
+ c \cos(2 \beta).
\ee
The phase $\beta$ adjusts itself such that $\cos (2 \beta) = - c / (4 b)$,
while $\alpha$ just follows it so that $a \cos (\alpha + \beta) = - |a|$.
SCPV manifests itself through the scalar-pseudoscalar mixing \cite{ma}
of the scalars $H^0$ and $R$ with the pseudoscalar $I$.
Indeed,
this mixing is indirectly generated.
If we write
\be
\sigma = e^{i \beta} \frac{u + \rho_3 + i \eta_3}{\sqrt{2}}\, ,
\ee
where $u \exp (i \beta) / \sqrt{2}$ is the VEV of $\sigma$,
then the mass matrix of the three scalars $H^0$,
$R$ and $\rho_3$,
and of the two pseudoscalars $\eta_3$ and $I$
is such that there is only one mass term connecting
the scalars with the pseudoscalars,
a term in $\rho_3 \eta_3$.
However,
this is enough to generate CP violation in the Higgs sector,
because the term $m (\Phi_1^\dagger \Phi_2 \sigma + h.c.)$
mixes $I$ with $\eta_3$.
This term is also crucial for generating a mass for $I$.

This Higgs potential is particularly interesting because
it allows us to give a naturalness argument
for the smallness of CP violation.
The soft-breaking parameter $\mu$ can naturally (in a technical sense)
be made small,
and therefore both $c$ in eq.~(\ref{V0}) and,
as a consequence,
the mass term mixing $\rho_3$ with $\eta_3$
are naturally small too \cite{nes}.
The potential is also interesting
because in it we see that,
though SCPV is generated exclusively in one part of the Higgs sector
(the complex singlet $\sigma$),
it is nevertheless fed into the relevant part of the Higgs sector
(the doublets $\Phi_1$ and $\Phi_2$).
In this respect,
this Higgs potential has some resemblance
with another potential suggested by one of us
\cite{lav} some time ago.

\paragraph{Different types of CP violation:}

We learn from the previous discussion
that there are four different possibilities for CP violation:

1) CP violation may be hard
but manifest itself only in the complexity of the CKM matrix $V$
(and therefore of the FCNI matrix $N_d$ too,
see eq.~(\ref{N})),
while the scalars $H^0$ and $R$
do not mix with the pseudoscalar $I$.
This is the situation in our first two examples.

2) CP may be broken in a hard fashion in the Yukawa couplings,
while simultaneously there are less phases in the VEVs
than terms seeing those phases in the Higgs potential.
In that case,
the CKM matrix is complex,
and besides the scalars $H^0$ and $R$ mix with the pseudoscalar $I$.
Our third and fourth examples would be scenarios for this,
irrespective of whether the couplings in the Higgs potential
are complex or real.

3) CP may be spontaneously broken.
Then,
the CKM matrix is real,
because of the special form of our Yukawa-coupling matrices
(see eqs.~(\ref{yu}), (\ref{yuc})).
The only CP-violating effect is the scalar-pseudoscalar mixing,
and the model becomes superweak-like in character.
This happens in our third and fourth examples
(assuming real Yukawa couplings and real constants in the Higgs potential).

4) The same as before,
but the horizontal symmetry $\S$ is softly broken
such that SCPV can be made arbitrarily small
in a technically natural sense.
This is the achievement of our fourth example.
\section{Constraints from $\epsilon$, $\Delta m_K$ and $\Delta m_B$}
The most severe restrictions on FCNI come from the $K^0 
\bar{K}^0$ and $B^0_d \bar{B}^0_d$ systems. In the following we will always
use $B^0$ as a shorthand notation for $B^0_d$. To calculate the 
contributions of the neutral Higgs bosons to $M_{12}$ we neglect $m_d$
compared to $m_s$ and $m_b$, use the vacuum insertion approximation
and obtain
\ba \nonumber
M_{12}^K (\varphi) & \approx & \frac{5}{24} \frac{f_K^2 m_K^3}{v^2}
                     (V_{1d}^* V_{1s})^2 \frac{1}{\M^2}\, , \\
M_{12}^B (\varphi) & \approx & \frac{5}{24} \frac{f_B^2 m_B^3}{v^2}
                     (V_{1d}^* V_{1b})^2 \frac{1}{\M^2}\, , \label{M12}
\ea
with
\be
\frac{1}{\M^2} \equiv \left( \frac{v_1}{v_2} + \frac{v_2}{v_1} \right)^2
                      \sum_k \frac{1}{M^2_k} \left( W_{2k}^2 - W_{3k}^2 +
                      2i W_{2k} W_{3k} \right).
\ee
For the numerical evaluation of eqs.~(\ref{M12}) 
we take $f_K = 112$ MeV,
$v = 246$ GeV,
$V_{us} = 0.22$ and $V_{cb} = 0.038$.
In the following we make use of the 
Wolfenstein parameterization \cite{wol} where the CKM matrix elements
are functions of the four parameters $\lambda$,
$A$,
$\rho$ and $\eta$.
Since we are only interested in orders of magnitude
we simply take $f_B = f_K$.

For the purpose of getting constraints on $\M$,
and estimating from them lower bounds on the scalar masses,
the relevant physical observables are
the CP-violating parameter
\be
|\epsilon| \approx \left|
\frac{\mbox{Im}\, (M_{12}^K e^{2i \xi_0})}
{\sqrt{2} \Delta m_K} \right|
\approx 2.3 \cdot 10^{-3},
\ee
the mass difference
of the $K^0 \bar{K}^0$ system
$\Delta m_K \approx 2 | M_{12}^K | \approx 3.5 \cdot 10^{-12}$ MeV
and the mass difference of the $B^0 \bar{B}^0$ system
$\Delta m_B \approx 2 | M_{12}^B | \approx 3.4 \cdot 10^{-10}$ MeV. 
In the phase convention inherent in the Wolfenstein parameterization
we can safely neglect the phase $\xi_0$ of the isospin-zero amplitude of the
two-pion decays of the kaons.
We require that the $\varphi$ contributions of eqs.~(\ref{M12})
to $\epsilon$, $\Delta m_K$ and $\Delta m_B$ do not exceed their respective
experimental values.
We thus get the following approximate restrictions
on $\M$:
\paragraph{Restrictions from $\epsilon$:}
\be \label{epsfit}
\begin{array}{lccl}
\mbox{u and c models:} & \left| \mbox{Im} \: \dfrac{1}{\M^2} \right| & < & 
4 \cdot 10^{-8} \: \mbox{GeV}^{-2} \\*[5mm]
\mbox{t model:} & \left| \mbox{Im} \: \dfrac{(1- \rho - i \eta )^2}{\M^2}
 \right| & < & 
2 \cdot 10^{-2} \: \mbox{GeV}^{-2}
\end{array}
\ee
\paragraph{Restrictions from $\Delta m_K$:}
\be \label{mKfit}
\begin{array}{lccl}
\mbox{u and c models:} & \dfrac{1}{|\M^2|} & < & 
7 \cdot 10^{-6} \: \mbox{GeV}^{-2} \\*[5mm]
\mbox{t model:} & \dfrac{(1- \rho)^2 + \eta^2}{|\M^2|} & < & 
3 \; \mbox{GeV}^{-2}
\end{array}
\ee
\paragraph{Restrictions from $\Delta m_B$:}
\be \label{mBfit}
\left. \begin{array}{lc}
\mbox{u model:} & \rho^2 + \eta^2 \\
\mbox{c model:} & 1 \\ 
\mbox{t model:} & (1-\rho)^2 + \eta^2
\end{array} \right\} \cdot \dfrac{1}{|\M^2|} < 4 \cdot 10^{-4} \: 
\mbox{GeV}^{-2}
\ee

Let us now look at what these restrictions
imply in various particular cases.

\paragraph{u and c models with HCPV:}
The u and
c models are quite similar.
If there is HCPV in the Yukawa couplings
but not in the $\varphi$ sector,
i.e.,
if $W_{2k} W_{3k} = 0$ for all $k$,
then $\Delta m_K$ gives a lower bound
on the neutral scalar masses $M_k$ of the order of a few hundred GeV.
If we have
in addition scalar-pseudoscalar mixing,
then the bound from $\epsilon$ applies,
and $M_k$ must be of order 1 to 10 TeV.

\paragraph{u and c models with SCPV:}
SCPV is more
interesting in the case of u and c models because, as we will see,
it leads to an effective superweak scenario.
In this case the corresponding
inequality in eq.~(\ref{epsfit}) must be replaced by the equality
\be \label{eps}
\left| \mbox{Im} \: \frac{1}{\M^2} \right|
= 4.4 \cdot 10^{-8} \: \mbox{GeV}^{-2},
\ee
because now the neutral scalars are the only source of CP violation.
If we have SCPV but no soft breaking of the horizontal symmetry $\S$,
we obtain again a lower bound of 1 to 10 TeV on $M_k$.
If we invoke soft breaking of $\S$,
we can argue that CP-violating amplitudes can be made naturally small
relative to CP-conserving ones.
We can then evade the tight restriction from $\epsilon$,
and the strongest bound on $M_k$
will be given by $\Delta m_K$.
In this way we again arrive at a
bound of a few hundred GeV for the scalar masses. 
We stress that these low masses 
are natural in the technical sense
in the case of the softly broken horizontal symmetry.
In both scenarios,
with or without softly broken horizontal symmetry,
the scalar masses are still too large
for the $\varphi$ contributions to $B^0 \bar{B}^0$ to be relevant.
Since the scalar contributions
to the two-pion decays of the kaons are minute anyway and,
as we will shortly see,
the electric dipole moment of the neutron (EDMN)
is also well below the experimental bound,
the u and c models with SCPV
constitute effective superweak models,
where CP violation can only be found in $\epsilon$.

We want to mention that the u and c models
would be distinguishable by the size of their effects
in the $B^0_s \bar{B}^0_s$ system.
But as these effects are negligible
the distinction between the u and c models is academic.

\paragraph{t models with SCPV:}
Here we want to show that the t models with SCPV are ruled out,
as was first pointed out by Joshipura \cite{josh}.
The reason is that $\mbox{Im}\, (1/\M^2)$ is fixed by $\epsilon$,
and one therefore gets a lower bound on $\Delta m_B$:
\be
\Delta m_B \geq 2 \sqrt{2} |\epsilon| \Delta m_K \left( \frac{f_B}{f_K}
\right)^2 \left( \frac{m_B}{m_K} \right)^3 \left( \frac{V_{tb}}{V_{ts}}
\right)^2.
\ee
This inequality is violated,
since experimentally the right-hand side
is more than one order of magnitude larger than the left-hand side.

\paragraph{t models with HCPV:}
Turning to HCPV,
we see that the relevant restriction on $M_k$ comes
from the $B^0 \bar{B}^0$ mass difference (see eq.~(\ref{mBfit})). 
$\epsilon$ poses no problem
in the case of HCPV because it can be fitted with the SM box diagram.
The model then allows for very light Higgs scalars,
with masses of order 100 GeV or even lower.
The effect on $\epsilon$ is very small,
but the structure of the total
$\Delta m_B$ is now of the form 
\be
\Delta m_B = [(1-\rho )^2 + \eta^2] | a(SM) + a(\varphi) |,
\ee
where $a(SM)$ is a real number,
which is derived from the SM box-diagram contribution.
Since the phase of the neutral-Higgs contribution $a(\varphi)$
is arbitrary if there is scalar-pseudoscalar mixing,
the neutral scalars have the effect of changing the radius
of the circle determined by $\Delta m_B$ in the $\rho$--$\eta$ plane.
In other words,
by varying the parameters of the Higgs sector,
we may shift the intersection area in the $\rho$--$\eta$ plane
allowed by  the constraints of $\epsilon$,
$V_{ub}/V_{cb}$ and $\Delta m_B$ \cite{pec}. 
If there is no scalar-pseudoscalar mixing,
then $a(\varphi)$ is real and,
at least in the vacuum insertion approximation,
it has the same sign as $a(SM)$,
thus shrinking the circle $(1-\rho)^2+\eta^2$.
On the other hand,
if there is scalar-pseudoscalar mixing,
$a(\varphi)$ has a phase relative to $a_{SM}$,
and then the predictions for CP violation
in the $B$-meson system get changed.

\paragraph{Electric dipole moments:}
In our models the one-loop electric dipole moments
of the u and d quarks (see for instance \cite{eck})
are of the order of magnitude
$e m_q^3/(16 \pi^2 M_k^2 v^2)$
($e$ is the electric charge of the positron and $q=u$ or $q=d$),
and therefore are well below $10^{-30}$ e$\cdot$cm.
In an oscillator model,
the exchange contributions to the EDMN
are one order of magnitude larger \cite{eck}
because one current quark mass factor $m_q$
gets replaced by a mass of order 300 MeV.

Let us however point out that the method of ref.~\cite{ans} gives,
when applied to the t models with CP violation
in the neutral-Higgs sector \cite{josh},
an EDMN of order of magnitude $10^{-25}$ e$\cdot$cm,
if the scalar masses are below 100 GeV,
which is very close to the experimental limit.
\section{Conclusions}
In this paper we have considered
a two-Higgs-doublet extension of the SM and,
within this framework,
we have studied a class of models
generated by a particular horizontal symmetry $\S$.
The main purpose of our investigation
was to reduce the inherent arbitrariness of two-Higgs-doublet models by
relating the FCNI couplings to the elements of the CKM matrix $V$.
With our choice of $\S$ we have achieved that the FCNI
among the down-type quarks are controlled by
factors $V_{\alpha i}^* V_{\alpha j}$,
while in the up sector there are no FCNI.
The only freedom left in the Higgs couplings
are the mixing angles of the neutral scalars
and the ratio $v_2 / v_1$ of the VEVs of the Higgs doublets.
The index $\alpha$ denotes one of the 
up-type quarks u, c or t,
and for each of these choices we have in principle a different model.
The second feature which distinguishes our
models is the type of CP violation.
As we have shown in section 3,
at least at the tree level it is possible to have a complex CKM matrix
but no CP violation in the neutral scalar sector,
or to have CP violation in both $V$
and the neutral-scalar mixings,
or to have SCPV and therefore a real CKM matrix.
In addition,
if we break the horizontal symmetry $\S$ softly,
SCPV can be made arbitrarily small by tuning the soft-breaking
term,
which is natural in the technical sense.
We want to stress once again that our form of the Yukawa couplings,
eqs.~(\ref{neutral}-\ref{N}),
has been obtained by invoking a horizontal symmetry
which is either spontaneously or softly broken.
Therefore,
no fine-tuning is involved
in the relationship between the CKM matrix elements and the FCNI.

Taking the above-mentioned possibilities into account we have 
altogether $3\times 4=12$ different scenarios.
However,
as we have shown,
the models with $\alpha =$ u and $\alpha =$ c are virtually indistinguishable,
and the t models with SCPV are ruled out.
This leaves us with six physically viable and different scenarios.

Among the u or c models the most interesting feature is the possibility
of SCPV,
thus returning to the original motivation for two-Higgs-doublet models
\cite{lee}.
However,
since the form of the Yukawa couplings is very constrained,
the predictive power of our models is much superior
to the one of the general two-Higgs-doublet model,
and we obtain the result that $\epsilon$ would be
the only observable CP-violating quantity.
We thus have an effective superweak model
with all CP violation residing in the scalar-pseudoscalar mixing.
If $\S$ is not softly broken,
but instead is spontaneously broken together with
the gauge group and CP,
the lower bound on the neutral-Higgs masses is of order 1 TeV or higher,
as in the generic two-Higgs-doublet model.
However,
if $\S$ is softly broken we can lower this 
bound to a few hundred GeV,
now enforced by $\Delta m_K$ instead of $\epsilon$.

In the case of the t models with HCPV the neutral Higgs masses can be as low
as 100 GeV,
this bound coming from $\Delta m_B$.
Here,
$\epsilon$ is reproduced by the SM box diagram,
while the scalar contributions in the $K^0 \bar{K}^0$ system
are negligible.
They do,
however,
contribute considerably in the $B^0 \bar{B}^0$ system and could
thus change the extraction of $|V_{td}|$ from $\Delta m_B$ and
alter the predictions of the SM for CP violation in this system.
Of course,
if CP violation is present in both $V$ and the neutral Higgs sector,
there would be more freedom to adapt our model to new experimental input.

We have thus shown that introducing a second Higgs doublet in the SM,
and possibly some gauge singlet scalars in the Higgs potential,
one can obtain interesting models with high predictive power.
On the one hand we have eradicated some of the disturbing features
of the generic two-Higgs-doublet model,
while on the other hand we have taken advantage of the new possibilities
offered by the scalar interactions.
\section*{Acknowledgements}
We wish to thank Gerhard Ecker and Jo\~ao P.\ Silva
for useful discussions and suggestions.
The work of G.\ C.\ B.\ was supported in part
by Science Project No.\ SCI-CT91-0729
and EC contract No.\ CHRX-CT93-0132.
%
%
%
%

%
%
\end{document}